\documentclass[12pt]{article}
\usepackage{bbm,latexsym}
\usepackage{amssymb,amsmath}
\usepackage{epsfig}
\newcommand{\be}{\begin{equation}}
\newcommand{\ee}{\end{equation}}
\newcommand{\ba}{\begin{eqnarray}}
\newcommand{\ea}{\end{eqnarray}}
\newcommand{\bs}{\begin{subequations}}
\newcommand{\es}{\end{subequations}}
\newcommand{\no}{\nonumber\\}
\newcommand{\mnu}{\mathcal{M}_\nu}
\newcommand{\mnuw}{\mathcal{M}_\nu^{(w)}}
\renewcommand{\sc}{\mathcal{S}}
\newcommand{\tc}{\mathcal{T}}
\newcommand{\diag}{\mbox{diag}}
\newcommand{\bone}{\mathbbm{1}}
\begin{document}
\renewcommand{\thefootnote}{\fnsymbol{footnote}}

\title{
\normalsize \hfill UWThPh-2012-24 \\
\normalsize \hfill CFTP/12-009 \\*[8mm]
\LARGE Maximal CP violation in lepton mixing \\
from a model with $\Delta(27)$ flavour symmetry}

\author{
P.M. Ferreira,$^{(1,2)}$\thanks{E-mail: ferreira@cii.fc.ul.pt} \
W.~Grimus,$^{(3)}$\thanks{E-mail: walter.grimus@univie.ac.at} \
L.~Lavoura,$^{(4)}$\thanks{E-mail: balio@cftp.ist.utl.pt} \
and P.O.~Ludl$ \, ^{(3)}$\thanks{E-mail: patrick.ludl@univie.ac.at}
\\[5mm]
$^{(1)} \! $
\small Instituto Superior de Engenharia de Lisboa \\
\small 1959-007 Lisbon, Portugal 
\\[2mm]
$^{(2)} \! $
\small Centre for Theoretical and Computational Physics,
University of Lisbon \\
\small 1649-003 Lisbon, Portugal
\\[2mm]
$^{(3)} \! $
\small University of Vienna, Faculty of Physics \\
\small Boltzmanngasse 5, A--1090 Vienna, Austria
\\[2mm]
$^{(4)} \! $
\small Technical University of Lisbon and CFTP \\
\small Instituto Superior T\'ecnico, 1049-001 Lisbon, Portugal
\\[2mm]
}

\date{8 August 2012}

\maketitle

\begin{abstract}
We propose a simple mechanism which enforces
$\left| U_{\mu j} \right| = \left| U_{\tau j} \right| \ \forall j =
1,2,3$ in the lepton mixing matrix $U$. 
This
implies
maximal atmospheric neutrino mixing
and
a maximal CP-violating
phase but does not constrain
the reactor mixing angle $\theta_{13}$.
We implement the proposed mechanism in two renormalizable
seesaw models which have features strongly resembling those of models
based on a flavour symmetry group $\Delta(27)$.
Among the predictions of the models, there is a determination, although
ambiguous, of the absolute neutrino mass scale, 
and a stringent correlation between the absolute neutrino
mass scale and the effective Majorana mass in neutrinoless double-beta
decay. 
\end{abstract}

\newpage

\renewcommand{\thefootnote}{\arabic{footnote}}

\section{Introduction}

With the recent results of the Double Chooz,
Daya Bay,
and RENO Collaborations~\cite{daya-reno}
the earlier hints~\cite{schwetz}
of a non-zero reactor mixing angle $\theta_{13}$ have been confirmed. 
The unexpectedly large value of $\theta_{13}$~\cite{daya-reno,tortola,fogli} 
renders a $\mu$--$\tau$ interchange symmetry
in the neutrino mass matrix~\cite{mu-tau-old},
and therefore also tri-bimaximal mixing~\cite{HPS},
highly unlikely,
since that symmetry was tailored to achieve $\theta_{13} = 0$
at some energy scale. 
However,
there is a different version of the $\mu$--$\tau$ interchange symmetry,
which is based on a generalized CP transformation
that includes the $\mu$--$\tau$ interchange~\cite{HScp,GLcp}.
In this version,
the maximal atmospheric mixing angle~$\theta_{23}$
is not coupled with a vanishing $\theta_{13}$
but rather with a maximal CP-violating phase $\delta$
in the lepton mixing matrix.
Phenomenologically,
this scenario is fully viable. 

In this letter we introduce a new mechanism
for generating this type of lepton mixing.
In order to establish our notation we
firstly define the lepton mass Lagrangian as
\begin{equation}
\mathcal{L}_\mathrm{mass} = 
- \bar\ell_L M_\ell \ell_R
+ \frac{1}{2}\, \nu_L^T C^{-1} \mnu \nu_L
+ \mbox{H.c.},
\end{equation}
with $M_\ell$ and $\mnu$ being the mass matrices
of the charged leptons and of the neutrinos,
respectively;
the latter mass matrix is of the Majorana type.
Those mass matrices are diagonalized
by $3 \times 3$ unitary matrices $U_\ell$ and $U_\nu$
according to
\bs
\ba
U_\ell^\dagger M_\ell M_\ell^\dagger U_\ell &=&
\mbox{diag} \left( m_e^2, m_\mu^2, m_\tau^2 \right),
\\
U_\nu^T \mnu U_\nu &=&
\mbox{diag} \left( m_1, m_2, m_3 \right),
\ea
\es
respectively.
Then the lepton mixing matrix $U$ is given by
\begin{equation}
U = U_\ell^\dagger U_\nu.
\end{equation}
Our idea is the following.
Suppose that we have a model which gives a \emph{real}\/ matrix $U_\nu$ and
\begin{equation}\label{Uomega}
U_\ell = U_\omega \equiv \frac{1}{\sqrt{3}} \left( \begin{array}{ccc}
1 & 1 & 1 \\ 1 & \omega & \omega^2 \\ 1 & \omega^2 & \omega
\end{array} \right),
\end{equation}
where $\omega = \left. \left( - 1 + i \sqrt{3} \right) \right/ 2$.
Then,
it is trivial to see that the mixing matrix $U$ has the property
\begin{equation}
\left| U_{\mu j} \right| = \left| U_{\tau j} \right|, \ \forall j = 1,2,3.
\label{wviot}
\end{equation}
When using the standard parameterization of the mixing matrix~\cite{rpp}, 
the relations~(\ref{wviot}) require~\cite{HScp,GLcp}
\bs
\ba
\cos\theta_{23} = \sin\theta_{23} &=& \frac{1}{\sqrt{2}},
\\
\sin \theta_{13} \cos \delta &=& 0,
\ea
\es
whence it follows
\bs
\ba
\theta_{23} &=& 
\frac{\pi}{4},
\\
\delta &=& \pm 
\frac{\pi}{2},
\ea
\label{viuto}
\es
since we know that $\theta_{13} \neq 0$.

This paper is organized as follows.
In section~\ref{model}
we develop two seesaw models based on the idea laid out above.
These
models make predictions beyond those in~(\ref{viuto});
the extra predictions are presented in section~\ref{predictions}.
The conclusions of the paper are summarized in section~\ref{concl}.

\section{The models}
\label{model}
The fermion sectors of our models
contain the usual leptonic Standard Model multiplets,
namely three left-handed gauge-$SU(2)$ doublets,
subsumed under the symbol $D_L$,
and three right-handed charged-lepton gauge singlets,
subsumed under the symbol $\ell_R$.
The scalar sectors contain three Higgs doublets with weak hypercharge $1/2$,
which we subsume under the symbol $\phi$.
For the symmetry transformations of the models we make use of the matrices 
\begin{equation}
E = \left( \begin{array}{ccc}
0 & 1 & 0 \\ 0 & 0 & 1 \\ 1 & 0 & 0
\end{array} \right),
\quad
A = \left( \begin{array}{ccc}
1 & 0 & 0 \\ 0 & -1 & 0 \\ 0 & 0 & -1
\end{array} \right),
\quad
C = \left( \begin{array}{ccc}
1 & 0 & 0 \\ 0 & \omega & 0 \\ 0 & 0 & \omega^2
\end{array} \right).
\end{equation}
We define the bilinears
\bs \label{phi}
\ba
\left[ \phi \ell_R\right]_0 &=& 
\phi_1 \ell_{1R} + \phi_2 \ell_{2R} + \phi_3 \ell_{3R},
\label{0} \\
\left[ \phi \ell_R\right]_1 &=& 
\phi_1 \ell_{1R} + \omega \phi_2 \ell_{2R} + \omega^2 \phi_3 \ell_{3R},
\label{1} \\
\left[ \phi \ell_R\right]_2 &=& 
\phi_1 \ell_{1R} + \omega^2 \phi_2 \ell_{2R} + \omega \phi_3 \ell_{3R},
\label{2}
\ea
\es
which are analogous to the ones used in  models
based on the symmetry $A_4$~\cite{rajasekaran}. 
Indeed,
under the transformations
\bs
\ba
\sc: & & \ell_R \to A \ell_R, \ \phi \to A \phi,
\\
\tc: & & \ell_R \to E \ell_R, \ \phi \to E \phi,
\label{tc}
\ea
\es
the bilinears transform as
\bs
\ba
\left[ \phi \ell_R\right]_j & \stackrel{\sc}{\longrightarrow} & 
\left[ \phi \ell_R\right]_j,
\\
\left[ \phi \ell_R\right]_j & \stackrel{\tc}{\longrightarrow} & 
\omega^{2j} \left[ \phi \ell_R\right]_j.
\ea
\es
These transformation properties allow us to write down
the charged-lepton Yukawa Lagrangian 
\begin{equation}\label{LYl}
\mathcal{L}_Y^{(\ell)} =
- \sum_{j=1}^3 h_j \bar D_{jL} \left[ \phi \ell_R \right]_{j-1}
+ \mbox{H.c.},
\end{equation}
if we
supplement~(\ref{tc}) by
\begin{equation}\label{tc1}
\tc: \quad D_L \to C^2 D_L.
\end{equation}
The charged-lepton Yukawa Lagrangian~(\ref{LYl})
looks very similar to the one
of some $A_4$-based models~\cite{rajasekaran,review},
but that look is misleading---in our
models
the roles of $D_L$ and $\ell_R$ are reversed
relative to
the $A_4$-based models.
This can also be seen by computing the mass matrix of the charged leptons,
which in our models is
\begin{equation}
\label{ml}
M_\ell = \diag \left( h_1, h_2, h_3 \right) 
\left( \sqrt{3} U_\omega \right) \diag \left( v_1, v_2, v_3 \right),
\end{equation}
where $v_j$ denotes the vacuum expectation value (VEV)
of the neutral component of the Higgs doublet $\phi_j$.
Thus,
$U_\omega^\dagger M_\ell$ is diagonal if $h_1 = h_2 = h_3$,
\textit{i.e.}\ if the Yukawa coupling constants are all equal,
whereas in the $A_4$ models one needs equality of VEVs. 
The equality of the $h_j$ is achieved by assuming
invariance of $\mathcal{L}_Y^{(\ell)}$ under
\begin{equation}
\label{tc'}
\tc': \quad D_L \to E D_L, \ \ell_R \to C \ell_R.
\end{equation}
Therefore,
in the following we shall use 
\bs
\ba
& h_1 = h_2 = h_3 \equiv
h, &
\\
\label{mll}
& U_\omega^\dagger M_\ell = \sqrt{3}\, h\, \diag \left( v_1, v_2, v_3 \right). &
\ea
\es
We emphasize that in our models we do \emph{not} make any assumption
of alignment of the VEVs of the Higgs doublets, 
\textit{viz.}\ we require neither any equality among the $v_j$
nor that any of them vanishes.
On the other hand,
we do require a strong hierarchy
of the VEVs; indeed, it follows from equation~(\ref{mll}) that
$m_e = \sqrt{3}\,\left| h v_1 \right|$, \textit{etc.}, and, therefore,
\be
|v_1| : |v_2| : |v_3| = m_e : m_\mu : m_\tau.
\ee

\subsection{Model I}

In this model we use the type~I seesaw mechanism~\cite{seesaw}.
The lepton sector contains three right-handed neutral gauge singlets $\nu_R$
and the scalar sector contains a fourth Higgs doublet,
$\phi_\nu$.
Additionally,
there are three \emph{complex}\/ scalar singlets,
which we subsume under the symbol $\eta$.
The Higgs doublet $\phi_\nu$ is invariant under $\sc$,
$\tc$,
and $\tc'$,
while $\nu_R$ and $\eta$ transform in the same way as $D_L$. 
A summary of the multiplets and their transformation properties
is presented in table~\ref{multiplets}.
\begin{table}[h]
\begin{center}
\begin{tabular}{|c|cccccc|}
\hline
symmetry & $D_L$ & $\ell_R$ & $\phi$ & $\nu_R$ & $\eta$ & $\phi_\nu$
\\ \hline 
$\sc$ & $\bone$ & $A$ & $A$ & $\bone$ & $\bone$ & 1
\\
$\tc$ & $C^2$ & $E$ & $E$ & $C^2$ & $C^2$ & 1
\\
$\tc'$ & $E$ & $C$ & $\bone$ & $E$ & $E$ & 1
\\ \hline
\end{tabular}
\caption{\label{multiplets} Multiplets of model~I
and their transformation properties.}
\end{center}
\end{table}
Notice that the symmetries $\tc$ and $\tc'$ together
generate a group $\Delta (27)$ under which $D_L$,
$\ell_R$,
$\nu_R$,
and $\eta$ are identical triplets while
$\phi$ decomposes into non-equivalent 
singlets.\footnote{We may speculate
about the full symmetry group of the model.
Taking into account also~$\sc$,
the column with caption~$\ell_R$ of table~\ref{multiplets}
suggests $AC$ to be a generator of the symmetry group;
it has the sixth root of unity,
$-\omega^2$,
in its diagonal.
So we might naively guess $\Delta (108) \equiv \Delta (3 \times 6^2)$
to be the symmetry group.
However,
beyond the symmetries listed in table~\ref{multiplets}
the model possesses
an accidental $2 \leftrightarrow 3$ interchange symmetry:
\be\nonumber
D_{2L} \leftrightarrow D_{3L}, \quad
\ell_{2R} \leftrightarrow \ell_{3R}, \quad
\phi_2 \leftrightarrow \phi_3, \quad
\nu_{2R} \leftrightarrow \nu_{3R}, \quad
\eta_2 \leftrightarrow \eta_3.
\ee
Therefore,
the full symmetry group is $\Delta(216) = \Delta (6 \times 6^2)$.
For a discussion of the groups $\Delta (6 n^2)$ see~\cite{luhn}.}

In this way we obtain the neutrino Yukawa couplings
\bs
\ba
\mathcal{L}_Y^{(\nu)} &=&
- y_\nu \bar D_L \tilde \phi_\nu \nu_R
\label{dirac} \\ & &
+ \frac{y}{2}\, \sum_{j=1}^3 \eta_j \nu_{jR}^T C^{-1} \nu_{jR}
\\ & &
+ y'\left( \nu_{2R}^T C^{-1} \nu_{3R} \eta_1
+ \nu_{3R}^T C^{-1} \nu_{1R} \eta_2 + \nu_{1R}^T C^{-1} \nu_{2R} \eta_3 
\right) + \mbox{H.c.}
\label{LYn}
\ea
\es
The neutrino Dirac mass matrix $M_D$,
which originates in~(\ref{dirac}),
is proportional to the unit matrix.
We assume the VEVs $\langle \eta_j \rangle_0 = s_j$
to be at a high (seesaw) scale. 
Therefore,
the \emph{inverse}\/ neutrino mass matrix has the form 
\begin{equation}\label{m-1}
\mnu^{-1} = \left( \begin{array}{ccc}
\zeta a & c & b \\ c & \zeta b & a \\ b & a & \zeta c
\end{array} \right),
\end{equation}
with $\zeta^\ast = y / y'$.
The inverse mass matrix~(\ref{m-1})
has the typical form of mass matrices
in renormalizable models based on the group $\Delta(27)$~\cite{ma}.
This is understandable since,
as we have pointed out,
the fields $D_L$,
$\nu_R$,
and $\eta$
behave under $\tc$ and $\tc'$
as irreducible three-dimensional representations of $\Delta(27)$.

As explained in the introduction,
we need the neutrino mass matrix $\mnu$ to be \emph{real}.
The first step in this direction is to assume in the Lagrangian
a CP symmetry which renders $h$,
$y_\nu$,
$y$,
and $y'$ real.
That symmetry is given by
\begin{equation}\label{cp}
\mbox{CP}: \quad 
\left\{ \begin{array}{l}
D_L \to iC D_L^*,\
\ell_R \to i S C \ell_R^*,\ 
\nu_R \to  iC \nu_R^*,
\\[1mm]
\phi \to S \phi^*,\
\phi_\nu \to \phi_\nu^*,\
\eta \to \eta^*,
\end{array} \right.
\end{equation}
where $C$ is the charge-conjugation matrix in Dirac space while
\begin{equation}
S = \left( \begin{array}{ccc} 1 & 0 & 0 \\ 0 & 0 & 1 \\ 0 & 1 & 0 
\end{array} \right)
\end{equation}
acts in flavour space.
This matrix $S$ is needed in order to interchange the terms
$\phi_2 \ell_{2R}$ and $\phi_3 \ell_{3R}$ in each of 
equations~(\ref{phi}).

Next we discuss the scalar potential of the three $\eta_j$.
It has six terms compatible with $\tc$,
$\tc'$,
and the CP symmetry:
\begin{eqnarray}\label{V}
V_\eta &=& \sum_{j=1}^3 \left( \mu \left| \eta_j \right|^2
+ \lambda_1 \left| \eta_j \right|^4 \right)
+ \lambda_2 \left(
\left| \eta_1 \eta_2 \right|^2
+ \left| \eta_1 \eta_3 \right|^2
+ \left| \eta_2 \eta_3 \right|^2
\right)
\no & &
+ M_1 \left( \eta_1 \eta_2 \eta_3 + \mbox{H.c.} \right)
+ M_2 \left( \eta_1^3 + \eta_2^3 + \eta_3^3 + \mbox{H.c.} \right)
\no & &
+ \lambda_3 \left( {\eta_1^*}^2 \eta_2 \eta_3
+ {\eta_2^*}^2 \eta_1 \eta_3
+ {\eta_3^*}^2 \eta_1 \eta_2 + \mbox{H.c.}
\right).
\label{pot}
\end{eqnarray}
All six constants in $V_\eta$ are real:
$\mu$,
$\lambda_1$,
and $\lambda_2$ are real because the potential is Hermitian
and $M_1$,
$M_2$,
and $\lambda_3$ are real because of the CP symmetry~(\ref{cp}).
The latter three terms in the potential
are responsible for the relative phases of the VEVs
$s_j$.
If we choose negative $M_1$,
$M_2$,
and $\lambda_3$,
then at the minimum of the potential those VEVs will have phases given
by~\cite{branco} 
\begin{equation}
\arg{s_j} = \omega^{p_j},
\label{phase}
\end{equation}
where the $p_j$ are integers such that $p_1 + p_2 + p_3 = 0\ \mathrm{mod}\, 3$.
This is precisely what is needed in order for the matrix $\mnu^{-1}$
to be real apart from unphysical phases.\footnote{Clearly,
there is a ninefold degeneracy of the minimum~(\ref{phase}),
hence there will be domain walls.
One could avoid them through an appropriate soft breaking,
for instance a term $-\mathrm{Re} \left( \eta_1 + \eta_2 + \eta_3 \right)$,
which breaks $\tc$ but conserves $\tc'$
and would render the case with real VEVs the deepest minimum.}

It can be shown that the potential~(\ref{pot}) is rich enough
to allow for the $\left| s_j \right|$ to be all different,
as needed in our model.

Since the model has four Higgs doublets,
one might be tempted to argue that,
after switching to a basis in the space of the Higgs doublets
where only one of them has a non-vanishing VEV, precisely
that doublet with VEV corresponds to the Higgs doublet of the Standard Model
and all other doublets can be made heavy~\cite{georgi}.
However,
this argument is only applicable in the general case
where the Higgs potential is not restricted by family symmetries.
But,
if one accepts the possibility of soft breaking
of the CP and family symmetries
through terms of dimension two in the scalar potential,
then one can apply the above argument in a modified way.
Ignoring,
for the time being,
the gauge singlets $\eta_j$,
we may assume that the term $\phi_\nu^\dagger \phi_\nu$ has \emph{negative} sign
while the $3 \times 3$ matrix of mass-squared terms for the $\phi_j$
is \emph{positive} definite.
Then,
the VEVs $v_j$ are induced by the VEV of $\phi_\nu$
through the soft-breaking terms $\phi_\nu^\dagger \phi_j$~\cite{GLR}. 
In this setting,
the role of the Higgs doublet of the Standard Model is played by $\phi_\nu$,
while the $\phi_j$ are additional doublets that can be made heavy.
When one includes in the scalar potential terms
containing products of both singlets $\eta$ and doublets $\phi$,
the potential becomes even more versatile with respect to our goal.

The models in this paper are designed for the lepton sector.
It is straightforward,
however,
to accommodate the quarks by using the doublet $\phi_\nu$ to give them masses,
thus playing the role of the Standard Model's sole Higgs doublet.
In this setting,
the doublets $\phi_j$ do not play any role in the quark sector.
A problem arises,
however,
since we have in the model a CP symmetry
and we know CP to be violated in the hadron sector.
This problem may be solved by adding to the model
one or more extra scalar doublets
transforming under the various symmetries
in exactly the same way as $\phi_\nu$;
all those doublets will have Yukawa couplings to the quarks and,
if their VEVs acquire relative phases through the mechanism 
of spontaneous CP violation,
then the quark mixing matrix will be complex,
yet the predictions for the lepton sector will stay unchanged,
because the Dirac mass matrix $M_D$ will just acquire an overall phase
which may be rotated away.

\subsection{Model II}

In this model we use the type~II seesaw mechanism~\cite{II}.
The lepton sector is identical with the one of the Standard Model,
\textit{i.e.}\ no right-handed neutrino singlets are present.
The scalar sector contains,
besides the three Higgs doublets in $\phi$,
three gauge-$SU(2)$ triplets with weak hypercharge $1$,
which we subsume under the symbol $\Delta$:
\be
\Delta = \left( \begin{array}{cc}
\Delta^+ \left/ \sqrt{2} \right. & \Delta^{++} \\
\Delta^0 & -\Delta^+ \left/ \sqrt{2} \right.
\end{array} \right).
\ee
We assume $\Delta$ to transform under $\sc$,
$\tc$,
and $\tc'$ in exactly the same way as $D_L$.
This allows us to write down the Yukawa couplings~\cite{kummer}
\bs
\ba
\mathcal{L}_Y^{(\Delta)} &=&
\frac{\tilde y}{2} \sum_{j=1}^3 D_{jL}^T C^{-1} \varepsilon \Delta_j D_{jL}
\\ & & + \tilde y' \left( D_{2L}^T C^{-1} \varepsilon \Delta_1 D_{3L}
+ D_{3L}^T C^{-1} \varepsilon \Delta_2 D_{1L}
+ D_{1L}^T C^{-1} \varepsilon \Delta_3 D_{2L} \right) + \mbox{H.c.},
\ea
\es
where 
\be
\varepsilon = \left( \begin{array}{cc} 0 & 1 \\ -1 & 0 \end{array} \right)
\ee
acts in
gauge-$SU(2)$ space.
When the neutral components of $\Delta$ acquire VEVs
$\left\langle \Delta_j^0 \right\rangle_0 = \delta_j$
we obtain
\be \label{m}
\mnu = \left( \begin{array}{ccc}
\tilde y \delta_1 & \tilde y' \delta_3 & \tilde y' \delta_2 \\
\tilde y' \delta_3 & \tilde y \delta_2 & \tilde y' \delta_1 \\
\tilde y' \delta_2 & \tilde y' \delta_1 & \tilde y \delta_3
\end{array} \right).
\ee
This is of the same form as the $\mnu^{-1}$ in equation~(\ref{m-1}).

From the transformation properties
of the scalar doublets $\phi$ and triplets
$\Delta$, it is obvious that
the scalar potential cannot have a trilinear term of the form 
$\phi^\dagger \Delta \varepsilon \phi^\ast$ 
invariant under $\sc$, $\tc$
and $\tc'$. Absence of such a term leads to a
Goldstone boson since
the scalar potential
becomes
invariant under separate phase transformations
of the $\phi$ and $\Delta$.
In order to avoid the Goldstone
boson one must resort
to soft breaking of the flavour symmetries.
We shall not pursue this issue further
and in the following we shall simply assume
that there is a satisfying solution which leads,
moreover,
to real VEVs at the minimum of the potential.

\section{Predictions for neutrino masses and lepton mixing}
\label{predictions}
We first discuss the predictions of model~I.
It is convenient to use the weak basis
where the charged-lepton mass matrix is diagonal.
In that basis the neutrino mass matrix is
\begin{equation}
\mnuw = U_\ell^T \mnu U_\ell.
\end{equation}
Since in our case $U_\ell = U_\omega$,
we find 
\begin{equation}
{\mnuw}^{-1} = U_\omega^\dagger \mnu^{-1} U_\omega^* =
\left( \begin{array}{ccc}
\bar\zeta \bar a & \bar c & \bar b \\
\bar c & \bar\zeta \bar b & \bar a \\
\bar b & \bar a & \bar\zeta \bar c
\end{array} \right),
\label{mnuw}
\end{equation}
where
\bs
\ba
\bar a &=& \frac{\zeta - 1}{3} \left( a + b + c \right),
\\
\bar b &=& \frac{\zeta - 1}{3} \left( a + \omega^2 b + \omega c \right),
\\
\bar c &=& \frac{\zeta - 1}{3} \left( a + \omega b + \omega^2 c \right),
\\
\bar \zeta &=& \frac{\zeta + 2}{\zeta - 1}.
\ea
\es
We know that $a$, $b$, $c$, and $\zeta$ are real. 
Therefore, $\bar a$ and $\bar \zeta$ are real too,
and $\bar c = \bar b^\ast$.
Therefore,
the inverse mass matrix~(\ref{mnuw}) has the form
\begin{equation}\label{mnuwf}
{\mnuw}^{-1} = 
\left( \begin{array}{ccc}
x & y & y^\ast \\ y & z & w \\ y^\ast & w & z^\ast
\end{array} \right),
\quad \mathrm{with} \; x \; \mathrm{and} \; w \; \mathrm{real}.
\end{equation}
This type of matrices was discussed in~\cite{GLcp},
where it was shown that its diagonalizing unitary matrix, 
which in the present case is the complex conjugate of the lepton mixing matrix,
\textit{i.e.}\ $U^*$,
is of the form~\cite{HScp,GLcp}
\begin{equation}\label{U}
U = \left( \begin{array}{ccc}
u_1 & u_2 & u_3 \\ w_1 & w_2 & w_3 \\ w_1^\ast & w_2^\ast & w_3^\ast
\end{array} \right),
\end{equation}
where the $u_j$ are real. 
Note that the phase convention inherent in equation~(\ref{U})
is that $U$ diagonalizes $\mnuw$
\emph{up to arbitrary signs}\/ of the masses.
Thus,
we have
\begin{equation}\label{mnuw1}
U^\dagger {\mnuw}^{-1} U^* = 
\diag \left( \frac{1}{\mu_1}, \frac{1}{\mu_2}, \frac{1}{\mu_3} \right),
\quad \mbox{with} \ \mu_j = \epsilon_j m_j, \ \epsilon_j = \pm 1.
\end{equation}
The phase factors appearing in the effective neutrino mass $m_{\beta\beta}$
of neutrinoless double-beta decay are identical
with the $\epsilon_j$,
\textit{i.e.}
\begin{equation}\label{mbb}
m_{\beta\beta} = \left| \sum_{j=1}^3 u_j^2 \mu_j \right| =
\left| \sum_{j=1}^3 u_j^2 \epsilon_j m_j \right|.
\end{equation}

There are nine physical quantities in the neutrino sector:
three neutrino masses,
three mixing angles,
one Dirac-type CP-violating phase,
and two Majorana phases. 
The mass matrix~(\ref{mnuw}) has four real parameters,
consequently the model must make five predictions:
the atmospheric mixing angle $\theta_{23}$ is $\pi / 4$,
the
Dirac-type
phase $\delta$ is $\pm \pi/2$,
each of the two Majorana phases is either $0$ or
$\pi$, and the
remaining prediction
can be gathered from inspection of equation~(\ref{mnuw}),
whence we deduce that
\begin{equation}\label{p5}
\left( {\mnuw}^{-1} \right)_{11}
\left( {\mnuw}^{-1} \right)_{13}
= \bar \zeta \bar a \bar b =
\left( {\mnuw}^{-1} \right)_{22}
\left( {\mnuw}^{-1} \right)_{23}.
\end{equation}
This identity is not meaningful in its phase,
because ${\mnuw}^{-1}$ may be rephased at will,
\begin{equation}
\left( {\mnuw}^{-1} \right)_{\alpha \beta} \to
\left( {\mnuw}^{-1} \right)_{\alpha \beta} e^{i \left( \psi_\alpha + \psi_\beta
  \right)}. 
\end{equation}
However,
the equality of the moduli of both sides of equation~(\ref{p5})
\emph{is}\/ physically meaningful.
The physical content of this relation will be worked out in the following.

Due to equations~(\ref{U}) and~(\ref{mnuw1})
one has
\begin{subequations}
\begin{eqnarray}
\left( {\mnuw}^{-1} \right)_{11} &=& \sum_{j=1}^3 \frac{u_j^2}{\mu_j},
\\
\left( {\mnuw}^{-1} \right)_{13} &=& \sum_{j=1}^3 \frac{u_j w_j^*}{\mu_j},
\\
\left( {\mnuw}^{-1} \right)_{22} &=& \sum_{j=1}^3 \frac{w_j^2}{\mu_j},
\\
\left( {\mnuw}^{-1} \right)_{23} &=& \sum_{j=1}^3 \frac{\left| w_j \right|^2}{\mu_j}. 
\end{eqnarray}
\end{subequations}
Notice that $\left( {\mnuw}^{-1} \right)_{11}$
and $\left( {\mnuw}^{-1} \right)_{23}$ are real,
\textit{cf.}~equation~(\ref{mnuwf}).
The equality of the moduli of both sides of equation~(\ref{p5}) then reads
\begin{equation}\label{inter}
\left( \sum_{j = 1}^3 \frac{u_j^2}{\mu_j} \right)^2
\left| \sum_{j^\prime = 1}^3 \frac{u_{j^\prime} w_{j^\prime}^*}{\mu_{j^\prime}}
\right|^2 
=
\left| \sum_{j = 1}^3 \frac{w_j^2}{\mu_j} \right|^2
\left( \sum_{j^\prime = 1}^3 \frac{\left| w_{j^\prime} \right|^2}
{\mu_{j^\prime}} \right)^2.
\end{equation}
At this point we have to exploit the unitarity condition of $U$
\be
\label{unitarity}
u_j u_{j^\prime} + 2\, \mathrm{Re} \left( w_j w_{j^\prime}^\ast \right)
= \delta_{j j^\prime}.
\ee
We can transform equation~(\ref{inter}) into
\begin{eqnarray}
& & \left( \sum_j \frac{U_j^2}{m_j^2}
+ \sum_{j < j^\prime} \frac{2 U_j U_{j^\prime}}{\mu_j \mu_{j^\prime}}
\right)
\left[
\sum_j \frac{U_j \left( 1 - U_j \right)}{2 m_j^2}
- \sum_{j < j^\prime} \frac{U_j U_{j^\prime}}{\mu_j \mu_{j^\prime}}
\right]
\no &=&
\left[ \sum_j \frac{\left( 1 - U_j \right)^2 }{4 m_j^2}
+ \sum_{j < j^\prime} \frac{- 1 + U_j + U_{j^\prime} + U_j U_{j^\prime}}
{2 \mu_j \mu_{j^\prime}} 
\right]
\no & & \times
\left[ \sum_j \frac{\left( 1 - U_j \right)^2}{4 m_j^2}
+ \sum_{j < j^\prime} \frac{1 - U_j - U_{j^\prime} + U_j U_{j^\prime}}
{2 \mu_j \mu_{j^\prime}} 
\right],
\label{result}
\end{eqnarray}
where we have defined $U_j \equiv u_j^2$.
Taking into account that the first line of $U$
is related to the mixing angles through
\bs
\label{u}
\ba
U_3 &=& \sin^2{\theta_{13}},
\\
U_2 &=& \cos^2{\theta_{13}} \sin^2{\theta_{12}},
\\
U_1 &=& \cos^2{\theta_{13}} \cos^2{\theta_{12}},
\ea
\es
the fifth prediction of our model,
embodied in equation~(\ref{result}),
is amenable to numerical analysis. 

In the case of model~II,
the predictions for lepton mixing
and for the Majorana phases,
listed in the paragraph after equation~(\ref{mbb}),
hold true as well.
As for relation~(\ref{result}),
we must make the replacement $\mu_j \to 1/\mu_j$
in order to obtain the corresponding relation for model~II.

\begin{figure}
\begin{center}
\epsfig{file=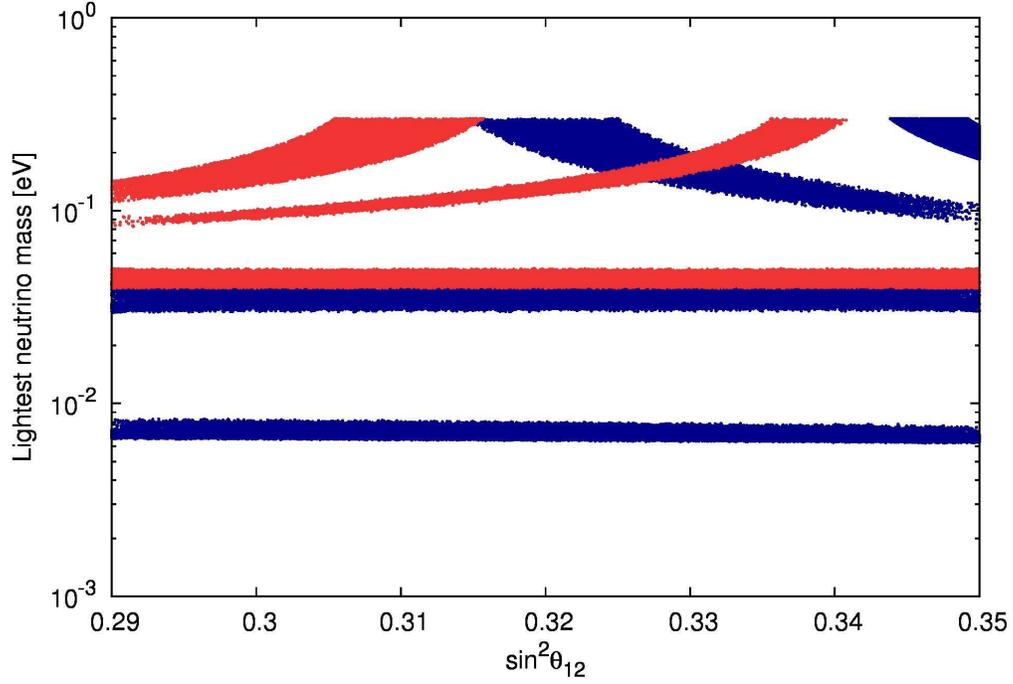,width=0.85\textwidth}
\end{center}
\vspace{-8mm}
\caption{Lightest neutrino mass as a function of 
  $\sin^2 \theta_{12}$ in the case of model~I. Here and in the
  following figures, dark (blue) 
  and light (red) colour indicate the allowed range for the normal
  and inverted ordering of the neutrino mass spectrum, respectively.
  \label{s12m0-1}} 
\end{figure}

\begin{figure}
\begin{center}
\epsfig{file=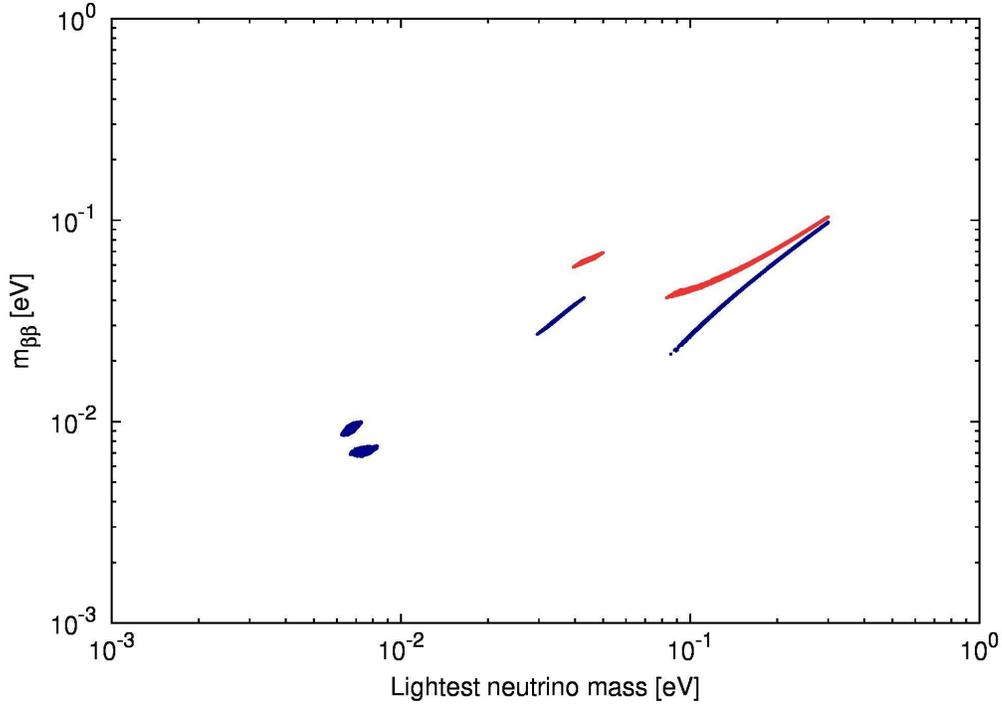,width=0.85\textwidth}
\end{center}
\caption{Effective Majorana mass $m_{\beta\beta}$ versus lightest
  neutrino mass in the case of model~I. \label{m0mbb-1}}
\end{figure}

\begin{figure}
\begin{center}
\epsfig{file=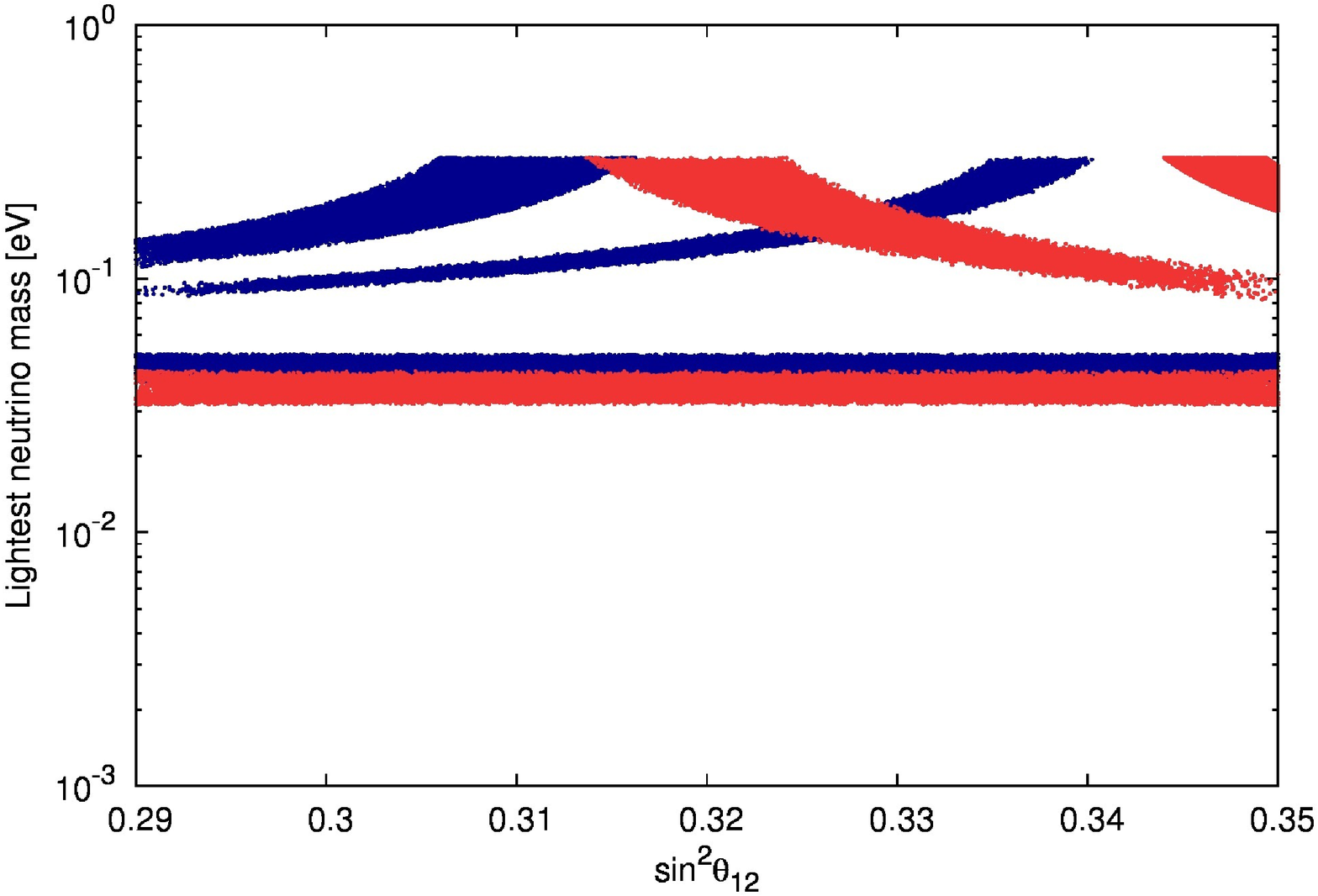,width=0.85\textwidth}
\end{center}
\caption{Lightest neutrino mass as a function of 
  $\sin^2 \theta_{12}$ in the case of model~II. \label{s12m0-2}}
\end{figure}

\begin{figure}
\begin{center}
\epsfig{file=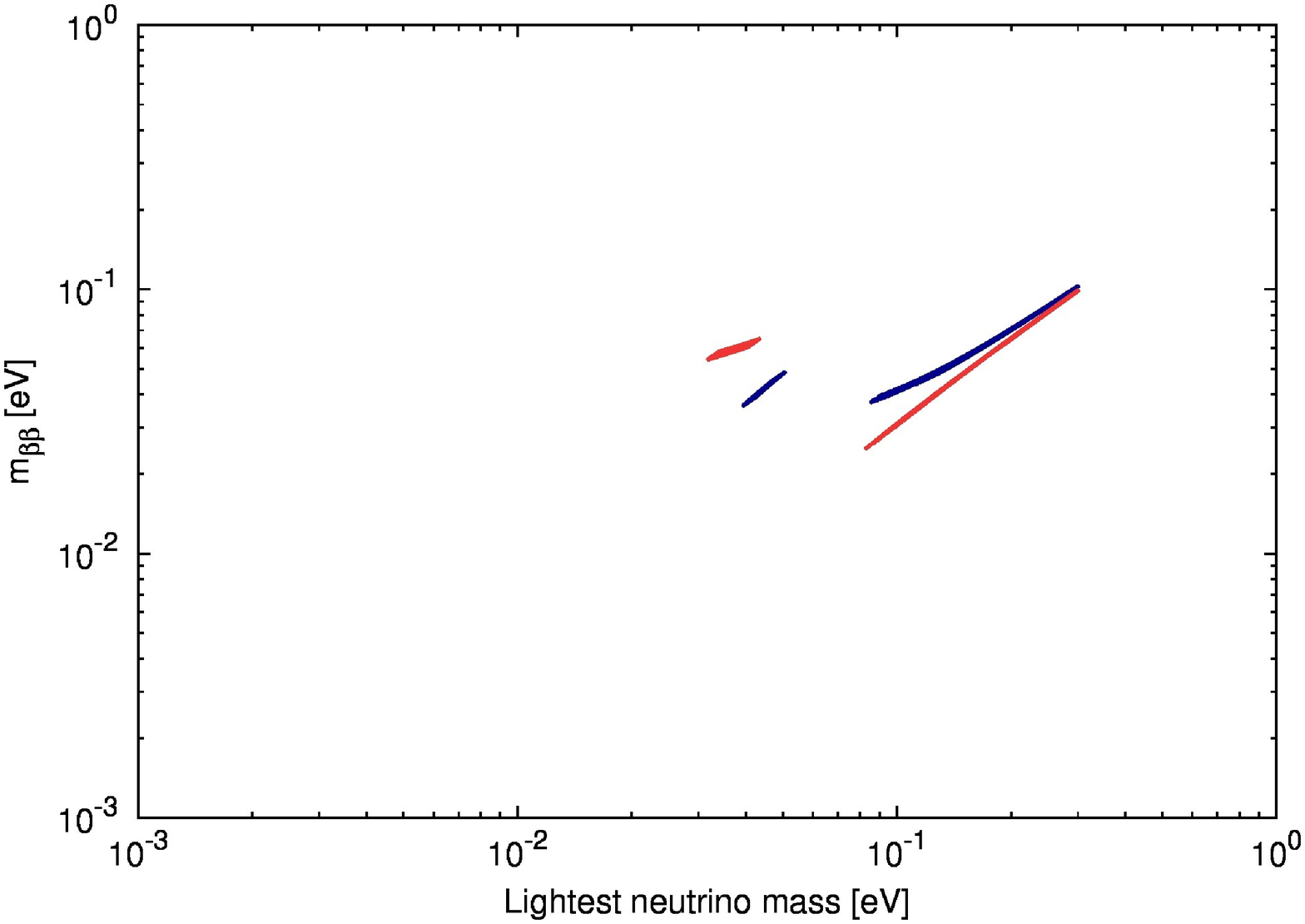,width=0.85\textwidth}
\end{center}
\caption{Effective Majorana mass $m_{\beta\beta}$ 
versus lightest neutrino mass in the case of model~II. \label{m0mbb-2}}
\end{figure}

In order to evaluate equation~(\ref{result}) numerically,
we use as input the 2$\sigma$ ranges of 
$\Delta m^2_{21} \equiv m_2^2 - m_1^2$,
$\Delta m^2_{31} \equiv m_3^2 - m_1^2$,
$\sin^2 \theta_{12}$,
and $\sin^2 \theta_{13}$ taken from table~1 of~\cite{tortola}. 
We allow the lightest neutrino mass,
$m_0$,\footnote{Note that $m_0 \equiv m_1$ for a normal neutrino mass spectrum
and $m_0 \equiv m_3$ for an inverted neutrino mass spectrum.}
to lie in between zero and $0.3\, \mbox{eV}$;
this is inspired by the extant cosmological bounds
on the sum of the light neutrino masses~\cite{rpp}.
These
five
parameters,
in the specified ranges,
form our parameter space.
In order to find points in this parameter space which are compatible
with equation~(\ref{result}),
we define a
figure-of-merit function
$F = \left. \left| R-L \right| \right/ | R+L | + \Phi$,
where $R$ and $L$ are the expressions
in the right-hand and left-hand sides,
respectively,
of equation~(\ref{result}),
and $\Phi$ is a function which has the value zero if all the parameters
lie in the ranges specified above
and $10^6$ if at least one parameter is outside its range.
We minimize $F$ and declare a point to be allowed whenever $F < 10^{-9}$.
For the minimization we employ the Nelder--Mead algorithm,
\textit{i.e.}\ the downhill simplex
method~\cite{nelder}.\footnote{This is the method
that we have used for producing the scatter plots displayed in this paper.
However,
it is also possible to treat equation~(\ref{result}) \emph{exactly},
since that equation produces,
when using as input the neutrino masses and $U_3$,
a quartic equation for $U_2$
(one must use $U_1 = 1 - U_2 - U_3$),
which is solvable through an exact algorithm.
We have used this exact method to confirm the numerical results
presented in this paper.}
In order to produce the scatter plots in figures~\ref{s12m0-1}--\ref{m0mbb-2}, 
for every possible sign choice\footnote{Note that the sign choices 
$(\epsilon_1,\epsilon_2,\epsilon_3)$
and $(-\epsilon_1,-\epsilon_2,-\epsilon_3)$
are equivalent,
as can be read off from equation~(\ref{result}).}
of the masses $\mu_j$---see
equation~(\ref{mnuw1})---we have run the Nelder--Mead algorithm
with $10^5$ randomly chosen simplices in the parameter space.

Figures~\ref{s12m0-1} and~\ref{m0mbb-1} are for model~I. 
In figure~\ref{s12m0-1} we have plotted the lightest neutrino mass $m_0$
as a function of $\sin^2 \theta_{12}$.
Here and in all other figures,
the blue (dark) colour corresponds to a normal
and the red (light) colour to an inverted neutrino mass spectrum.
The different bands in figure~\ref{s12m0-1} 
are associated with different sign choices $\epsilon_j$.
These bands correspond to the allowed spots and lines in figure~\ref{m0mbb-1},
which can be deduced from
the corresponding ranges of the smallest neutrino mass.
Let us consider figure~\ref{m0mbb-1}
for a detailed explanation of the sign choices
associated with the allowed ranges of $m_0$.
Of the two spots in the left of that figure,
the upper one corresponds to $(+++)$ and the lower one to $(++-)$.
In the middle of the scatter plot,
both the upper and the lower stroke correspond to signs $(++-)$.
On the right part of the figure,
the upper line corresponds to both $(+-+)$ and 
$(+--)$ but the second sign choice holds only in its upper three quarters;
the lower line,
too,
is generated by two sign choices:
$(+--)$ holds along the whole line and $(+-+)$ in its upper half.

Figures~\ref{s12m0-2} and~\ref{m0mbb-2} refer to model~II.
The major difference between models~I and~II is that
the two lowest bands for $m_0$ disappear in model~II:
there is no band below $m_0 = 10^{-2}$\,eV in figure~\ref{s12m0-2}
and there are no spots in the left side of figure~\ref{m0mbb-2}.
This also means that the sign choice $(+++)$ is not allowed in model~II.
Otherwise,
the interpretation with respect to the signs $\epsilon_j$
is the same in both models.

In both models,
the dependence of the allowed range of $m_0$ on
$\sin^2{\theta_{13}}$
is very faint and does not show up significantly in a plot.
For this reason we refrain from showing those plots here.

\section{Conclusions}
\label{concl}
The recent experimental results on neutrino oscillations have shown that
the reactor mixing angle $\theta_{13}$ is not as small as previously thought.
This disagrees with the standard version
of $\mu$--$\tau$ interchange symmetry,
but not with an alternative version
which predicts a maximal atmospheric mixing angle
and maximal CP violation in neutrino mixing
while leaving $\theta_{13}$ arbitrary.
In this paper we have introduced a novel mechanism
which realizes this scenario.
Our mechanism needs
a left-handed diagonalization matrix $U_\ell = U_\omega$
in the charged-lepton sector---see equation~(\ref{Uomega})---where $U_\omega$
is the well-known maximal-mixing unitary matrix
which also appears in ordinary $A_4$-based models;
our mechanism moreover needs
a real neutrino mass matrix.
We have constructed two models,
one based on the type~I and the other one based
on the type~II seesaw mechanism.
Since with regard to the neutrino sector
the symmetry structure of our models bears resemblance
with models based on $\Delta(27)$,
we have obtained an additional constraint on the neutrino mass
and mixing parameters,
which can be approximately interpreted
as a determination of the absolute mass scale
of the neutrinos in terms of the mass-squared differences
and of the solar mixing angle---see figures~1 and~3.
Because several sign choices are possible for the neutrino 
masses---see equation~(\ref{mnuw1})---this determination is,
however,
ambiguous.
Since in this alternative version of $\mu$--$\tau$ interchange symmetry
the Majorana phases are either zero or $\pi$,
there is a rather stringent correlation
between the neutrino mass scale and $m_{\beta\beta}$,
the effective Majorana mass in 
neutrinoless double-beta decay---see figures~2 and~4.
It should be emphasized that $m_{\beta \beta}$ may in our model
assume quite large values,
of order $0.1\, \mbox{eV}$,
which might render it observable
in upcoming experiments---see~\cite{faessler} and references therein.

Besides the predictions for the lepton sector,
the models have interesting features.
They do not require any VEV alignment, only reality of the VEVs
of the scalar gauge singlets in model~I / gauge triplets in model~II
is essential.
This reality is easily obtained at least in the case of model~I,
as we have shown here and also earlier in~\cite{realvevs}. 
Possible phases of the VEVs of the scalar doublets
are irrelevant for CP violation in our models.
Nevertheless,
the models do feature CP violation in lepton mixing, 
which is indeed maximal,
\textit{i.e.}\ $\delta = \pm \pi/2$.
The reason is that here the CP symmetry~(\ref{cp})
is not broken by complex VEVs
but rather by $\left| v_2 \right| \neq \left| v_3 \right|$,
which is responsible for the muon and tau masses being different, 
($m_\mu \neq m_\tau$)~\cite{GLcp,mcp}. 
Indeed,
in our models the different masses of the charged leptons
are not obtained from different Yukawa coupling constants
but rather from different VEVs.
This idea has been used before in several other models~\cite{haber,hps,zee};
here we have followed~\cite{hps} in this respect.

\vspace{3mm}

\noindent
\textbf{Acknowledgements:} 
The work of P.M.F.\ is supported by the Portuguese
Foundation for Science and Technology (FCT)
through the contract PTDC/FIS/117951/2010,
the FP7 Reintegration Grant PERG08-GA-2010-227025,
and PEst.OE/FIS/UI0618/2011. 
The work of L.L.\ is funded by FCT
through its unit 777 and through the project PTDC/FIS/
098188/2008. W.G.\ and P.O.L.  acknowledge support from the Austrian
Science Fund (FWF), Project Nr.\ P~24161-N16.

\end{document}